\documentclass[letterpaper, 10 pt, conference]{ieeeconf}  

\IEEEoverridecommandlockouts                              

\usepackage{icarmacros}

\title{Towards Live Programming in ROS with PhaROS and LRP}
\author{Pablo Estefó$^{1}$, Miguel Campusano$^{2}$, Luc Fabresse$^{1}$,\\Johan Fabry$^{2}$, Jannik Laval$^{1}$, and Noury Bouraqadi$^{1}$
\thanks{$^{1}$Dépt. IA - Ecole des Mines de Douai
        {\tt\small {firstname.lastname@mines-douai.fr}
        \url{http://car.mines-douai.fr}}}%
\thanks{$^{2}$PLEIAD and RyCH labs, Computer Science Department (DCC), University of Chile, Chile
}%
}

\begin{document}
\maketitle
\thispagestyle{empty}
\pagestyle{empty}

\begin{abstract}
In traditional robot behavior programming, the edit-compile-simulate-deploy-run cycle creates a large mental disconnect between program creation and eventual robot behavior. This significantly slows down behavior development because there is no immediate mental connection between the program and the resulting behavior. With live programming the development cycle is made extremely tight, realizing such an immediate connection. In our work on programming of ROS robots in a more dynamic fashion through PhaROS, we have experimented with the use of the Live Robot Programming language. This has given rise to a number of requirements for such live programming of robots. In this text we introduce these requirements and illustrate them using an example robot behavior.
\end{abstract}

\section{Introduction}
\label{sec:intro}

In Live Programming~\cite{tanimoto90}, the development cycle is made extremely tight: program edits are continuously integrated in the always-running program and their effects are immediately visible. As a result, there is no cognitive dissociation between writing the code and observing its execution, and hence programmers have an immediate connection with the program they are making.

In this paper, we present early experiments introducing live programming into the development cycle of robotic applications. We target the ROS~\cite{rosICRA09} middleware because it is a de-facto standard that moreover has an availability of numerous packages provided by an active community backed by the Open Robotics Software Foundation.

However, ROS does not provide any support for live programming.
The development cycle of ROS introduces a clear cut between the edit, compile, and run steps of software development.
This division already appears in its core concepts since a ROS package is a static entity, whereas a ROS node is a run-time entity.

Nodes to run as well as their name spaces and parameters are statically expressed in a {\sf .launch} file and cannot be changed at run-time.
So, the whole architecture of a ROS-based application is static and, traditionally, any small change requires restarting the entire application.

As a first step in exploring live programming with ROS, we combined the Live Robot Programming (LRP) language~\cite{lrpdef}: a high-level DSL for live programming of robots, with PhaROS~\cite{pharosWeb}: a bridge between ROS and the dynamic language Pharo~\cite{Berg13a,pharoByExample1}. This work served as the foundation for the integration of ROS in LRP. As a result of these experiments, we have encountered a number of requirements for live programming with LRP and ROS, and we present them in this text.

\section{Requirements for Live Programming Robots with ROS}
\label{sec:requirements}

A ROS application is a graph of nodes connected by topics.
Enabling liveness in ROS implies making it possible to dynamically change any part of the graph.
This means :

\paragraph*{Requirement 1} Starting and stopping individual nodes or subsets of nodes as many times as required during the application lifetime.
For instance, when two nodes are communicating through a topic and the subscriber is shut down, when relaunched it should reconnect and resume communication.

\paragraph*{Requirement 2} Changing node parameters at run-time after the node has been launched. 

\paragraph*{Requirement 3} Partially or fully replacing node code at run-time without stopping it. 

\paragraph*{Requirement 4} Changing node connections at run-time (\ie~the edges of the graph). For any given node, we should be able to dynamically change topics to which it subscribes or publishes.

\section{Proposed Solution: LRP+PhaROS}
\label{sec:contribution}

Our solution is implemented using the Pharo dynamic language.
It consists of the LRP DSL combined with PhaROS: a ROS client for Pharo.

\subsection{LRP}
Live Robot Programming (LRP)~\cite{lrpdef} is a live programming language, of nested state machines, together with an interpreter and visualization (all of which is implemented in Pharo). This kind of machines are said to map well to the problem domain, which is corroborated by the fact that multiple Robocup teams have performed outstandingly in the football competition using these languages~\cite{loetzsch06xabsl, kouretes12}.

The main differentiator of LRP with regard to other languages using the nested state machine paradigm is its nature as a live programming language. Because of this, LRP code is \emph{always} running and LRP has a well-defined semantics for how it behaves in the face of code that is syntactically malformed, incomplete or erroneous programs, and changes to the code while a program is running. Put briefly, as a live programming language, the priority is to keep the program running in spite of a wide variety of errors. Also, program changes are integrated while the code runs, only restarting program execution when this integration cannot be performed. A full discussion on these features and the language in general is however outside of the scope of this paper. For more information we refer to published work~\cite{lrpdef}, as well as the LRP website\footnote{Website URL: \url{http://pleiad.cl/lrp}}.

LRP is designed for the programming of robot behaviors, and the communication between LRP and the actual robot is through the use of APIs of specific robot platforms or middleware. The integration of ROS in LRP~\cite{lrpdef} was achieved based on the results of the experiments we report here.

The LRP language is built based on the following concepts:
\begin{itemize}
\item \ct{Machines} with \ct{states} and different kinds of \ct{transitions}. 
\item \ct{Transitions} that occur on \ct{events}, timeouts or immediately after entering in a \ct{state}.
\item \ct{Action blocks} are snippets of Smalltalk code that have access to the complete Smalltalk environment, as well as LRP variables.
\item \ct{Variables} can be defined in machines, are initialized when declared using an action block, and are lexically scoped.
\item \ct{Events} are explicitly defined, triggering if their action block evaluates to true, \ie~any piece of Smaltalk code can serve as a guard for an event.
\item \ct{States} can have action blocks that are run when entering, leaving or when they are active, again enabling any piece of Smalltalk code to be used.
\item States can also define state machines, which enables nesting.
\end{itemize}

\subsection{PhaROS}
PhaROS \cite{pharosWeb} provides a framework and a set of tools for developing ROS nodes in the Pharo language and environment. 
Pharo is a pure object-oriented programming language coupled with a dynamic environment.

In PhaROS, ROS nodes are reified as objects, allowing the developer to build and control their execution at a higher level of abstraction. 
PhaROS also contains tools allowing the deployment of catkin packages: It automatizes the generation of xml launch files, makefiles, type and scripts creation. The intent of this is to let the programmer focus on programming and not on creation and maintenance of infrastructure.

PhaROS nodes can be restarted as many times as needed, due to the dynamic nature of the Pharo environment. Instances of ROS nodes can be terminated at any time and are garbage collected. New instances can be launched directly from Pharos without a need to restart ROS as they transparently reconnect with external existing ROS topics or parameters.

\subsection{Integration of LRP and PhaROS}
As said above, LRP relies on the existence of an API to the robot (middleware) being used. 
In our experiments we worked with a Robulab robot that runs ROS and therefore we needed to construct such an API.
 
\begin{figure}[t]
   \begin{center}
   \includegraphics[scale=0.48]{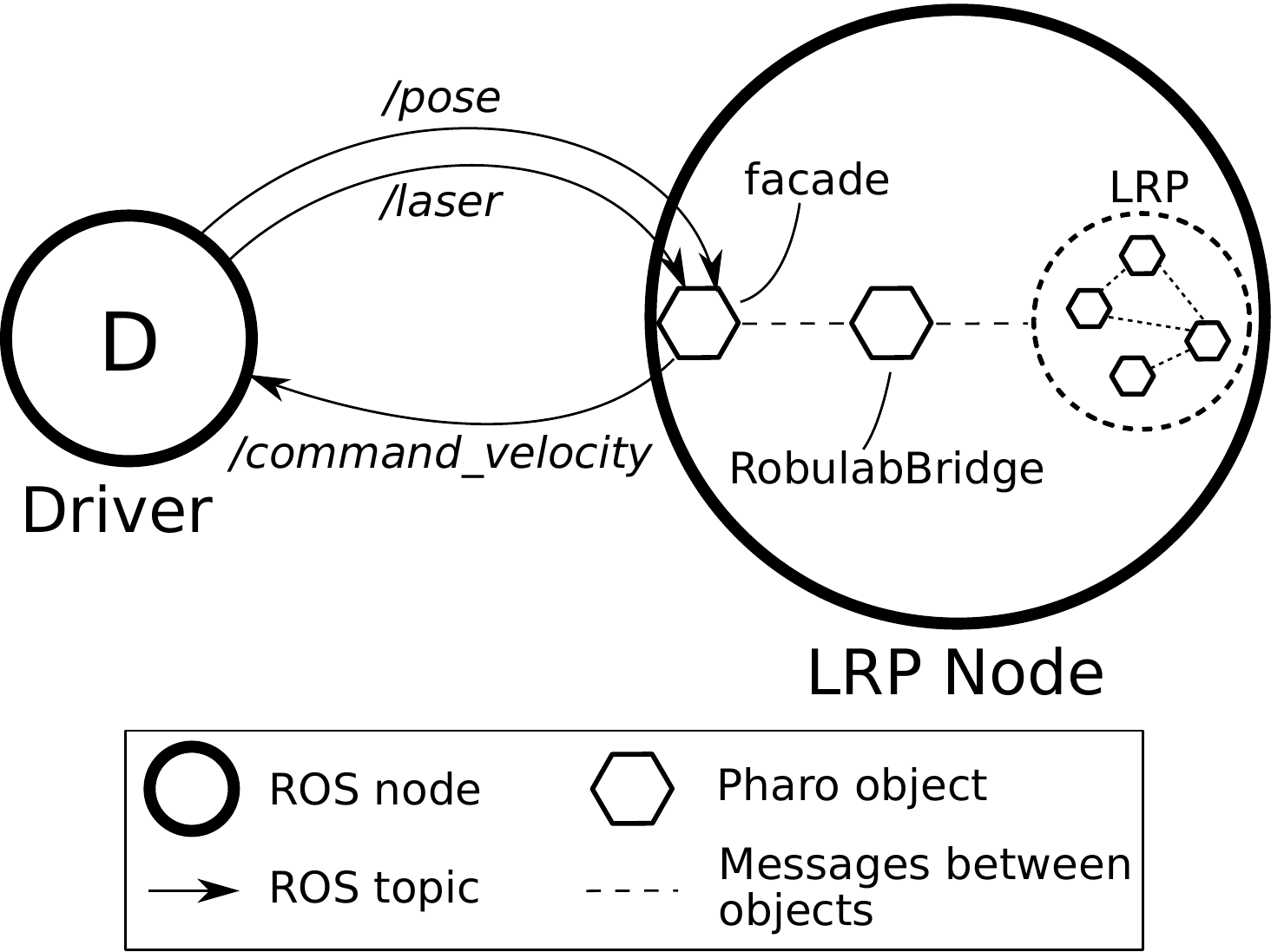}
   \caption{ROS graph of application shown in experiment.}
   \label{fig:diagram-ros}
   \end{center}
\end{figure}

To do this, we implemented a bridge module to ROS as well as a specific interface for the Robulab. The former takes the role of a facade class (\ct{facade} in \figref{diagram-ros}). It provides access to external ROS resources such as topics or parameters. The latter, a \ct{RobulabBridge} class, is tailored to the Robulab and provides an API for the features specific to the robot, hiding the particularities of ROS. 
Note that this abstraction of specifying an API specific to the features of the robot can be generalized to any ROS node, where the facade could be simply reused.

The robot's node (\ct{Driver} in \figref{diagram-ros}) consumes messages on the \ct{/command\_velocity} topic and publishes messages on the \ct{/laser} and \ct{/pose} topics. 
Hence, the \ct{RobulabBridge} provides methods that wrap messages to and from these topics such as:

\begin{itemize}
\item {\bf\ct{forward: linearSpeed}} publishes on the \ct{/command\_velocity} topic to make the robot move linearly forward at a \ct{linearSpeed} speed.
\item  {\bf\ct{turn: angularSpeed}} similar to \ct{forward:}, making the robot rotate at \ct{angularSpeed}.
\item  {\bf\ct{isThereARightObstacle: minimumDistance}} selects laser data corresponding to the front-right part of the laser beam, and checks if there is an obstacle at a distance less or equal than \ct{minimumDistance}.
\end{itemize} 

\section{Experiment: writing an obstacle avoider}
\label{sec:experiment}

In this section, we report on an experiment written in LRP that runs on top of PhaROS and therefore uses ROS. 
The example we present is simple, but interesting nonetheless because it exemplifies what is possible with live programming. Videos of the experience are available on \url{http://car.mines-douai.fr/} . 

In the experiment, we will change a mobile robot's obstacle avoidance behavior incrementally at runtime. 
First we will implement a simple solution that lets the robot move without crashing into obstacles.
After that the behavior will be evolved by adding obstacle avoidance.
All of this will be done without needing to restart any part of the ROS platform or equipment used, all software changes are immediately perceived as robot behavior changes. 

The experiment was performed using a Robulab robot which can be translated and/or rotated on the floor. The robot is equipped with a Sick S300 laser range sensor, which detects obstacles up to 30 meters within 270 degrees around the robot. Technical specifications of the robot are available on its website\footnote{Website URL: \url{http://www.robosoft.com/products/indoor-mobile-robots/robulab/robulab-10.html}}.

\subsection{Stop when an obstacle is detected}
\label{sec:stop-forward}

First, we will define the four variables that we will use during the experiment. 
Listing~\ref{codeLrp1} shows how they all are defined. 
\ct{f\_vel} and \ct{t\_vel} are both linear and angular velocities for the robot respectively (set to 0.25 $[m/s]$ and 0.5 $[rad/s]$ respectively) and \ct{min\_distance} (set to 0.5 $[m]$) is the minimum distance to a physical object to be considered as an obstacle.
The fourth variable is initialized with a reference to an instance of \ct{RobulabBridge} class, presented before. 
Note that \ct{t\_vel} it is not used in the first behavior we implement, but it could be added later at runtime.

\begin{code}[caption={Initialization of the live programming session},label=codeLrp1]
(var f_vel := [0.25])
(var t_vel := [0.5])
(var min_distance := [0.5])
(var robulab := [RobulabBridge uniqueInstance])
\end{code}        

The first behavior to describe in LRP is to make the robot move forward and stop when an obstacle is detected in front.
So a simple program should consider two states: \ct{forward} and \ct{stop} (lines 2 to 5 in Listing~\ref{codeLrp2}). 
When entering the forward state, due to the statement in line 3 the program will send the message \ct{forward:} with \ct{f\_vel} as argument to the \ct{robulab} variable. 

As a result the robot will move forward with a speed of \ct{f\_vel} $[m/s]$. 
When the robot detects that there is an obstacle at distance of \ct{min\_distance} $[m]$ or closer, the \ct{obstacle} event will occur, as specified in line 8, and the machine will perform a state change from \ct{forward} to \ct{stop} through the \ct{t-stop} transition (line 6).
When entering the \ct{stop} state (line 5) the message \ct{stop} is sent to the \ct{robulab} making the robot stop moving. 
If suddenly the obstacle disappears, the event \ct{notObstacle} occurs (lines 10-11), and through the \ct{t-forward} transition, the machine changes to \ct{forward} state making the robot move again.

The last line of Listing~\ref{codeLrp2} initializes the state machine in the \ct{forward} state.

\begin{code}[caption={First behavior of state machine describing its states, transitions and events in LRP},label=codeLrp2]
(machine Tito
    ( state forward
        ( onentry [robulab forward: f_vel] ))
    ( state stop
        ( onentry [robulab stop] ))
    (on obstacle forward -> stop t-stop)
    (on noObstacle stop -> forward t-forward)
    (event obstacle 
        [robulab isThereAnObstacle: min_distance])
    (event noObstacle 
        [(robulab isThereAnObstacle: min_distance) not ])       
)
(spawn Tito forward)
\end{code}

\subsection{Avoiding obstacles}
\label{sec:avoiding-obstacles}

\begin{code}[caption={Aditional code for including obstacle avoidance behavior},label=codeLrp3]   
    ( state turnLeft
      ( onentry [robulab turn: t_vel] ))
    ( state turnRight
      ( onentry [robulab turn: t_vel negated] ))         
    (on rightObstacle stop -> turnLeft t-lturn)
    (on leftObstacle stop -> turnRight t-rturn)
    (on noObstacle turnLeft -> stop t-tlstop)
    (on noObstacle turnRight -> stop t-trstop)   
    (event rightObstacle [robulab isThereARightObstacle: min_distance])
    (event leftObstacle [robulab isThereALeftObstacle: min_distance]) 
\end{code}

Now we have the robot stopped in front of an obstacle, so the current status is \ct{stop}.
Next, a simple obstacle avoidance behaviour is added: the robot will turn left or right when an obstacle is detected and when there is no obstacle it will move forward, as already defined.

We define two more states: \ct{turnLeft} and \ct{turnRight} (lines 1 to 4 in Listing~\ref{codeLrp3}). 
They are reached by a simple obstacle detection algorithm, and in them the message \ct{turn:} is sent to \ct{robulab} with the same speed (\ct{t\_turn}) but a different direction depending on the state.
Considering the obstacle detection algorithm, the \ct{RobulabBridge} provides useful methods to know if the obstacle is in the left or the right side of the front of the robot. 
These are used to emit \ct{rightObstacle} or \ct{leftObstacle} events when needed (lines 9 through 10).
These events make the machine change from \ct{stop} to the turning states by the \ct{t-tlturn} and \ct{t-trturn} transitions (lines 5 and 6). 
When no obstacle is detected, \ct{noObstacle} is raised and machine changes the state to \ct{stop}, via the transitions of lines 7 and 8, which will be immediately followed to a transition to \ct{forward} via \ct{t-forward}.

When writing this code, immediately when the lines of codes defining the \ct{rightObstacle} or \ct{leftObstacle} events are added, one of these occur (since the robot is stopped at an obstacle). 
This then makes the robot turn to avoid the obstacle, and move forward when the obstacle is avoided.

\section{Requirements evaluation}
In this section, the fulfilment of requirements listed in \secref{requirements} is discussed.

\paragraph*{Requirement 1} This requirement is fulfilled for PhaROS nodes.
In PhaROS nodes are objects with specific methods for both initializing and finalization.
Once they are finalized, they do not continue working nor interacting with the rest of the ROS platform.
Nodes can be restarted multiple times, reconnecting to necessary topics and start publishing immediately. 

\paragraph*{Requirement 2} Typically, node parameters are statically defined in a \ct{.launch} file and once launched, they are cannot be changed. 
In contrast, in PhaROS a ROS node is an object and their parameters are materialized as fields accesible through methods. As a consequence, they can be changed at run-time either by any piece of Smalltalk code, for example inside an action block of an LRP program.
Furthermore, parameters can also be changed at startup when using the \ct{rosrun} command using command-line options for changing parameters (\ct{rosrun package node \_parameter=value}).

\paragraph*{Requirement 3} The logic behind the use of data from subscriptions, its processing and publishing can be modified when the PhaROS node was already launched. 
This is illustrated in detail in \secref{experiment} as the response of the robot to obstacles evolved at runtime. More specifically, the developer tuned robot behavior in an iterative development process to get the expected response.  

\paragraph*{Requirement 4}
By default in ROS, node connection relies on topic names that are hardwired inside the code of nodes. 
Still, at deployment time this can be altered using namespaces. 
Changing node connections at run-time means changing topic names and namespaces at run-time.
PhaROS provides runtime subscription/publishing which implies that topics are created when the ROS node is being executed.

\section{Related Work}
\label{sec:related}

To the best of our knowledge, LRP is the only work that proposes the live programming of robot behaviors through nested state machines. Live programming was originally proposed by Tanimoto on Viva~\cite{tanimoto90}, a visual programming language for image manipulation. We are aware of two DSLs for robot behavior programming based on nested state machines: the Kouretes Statechart Editor~\cite{kouretes12} and XABSL~\cite{loetzsch06xabsl}.

\section{Conclusion}
\label{sec:conclusion}

In this work we have presented four requirements to support live programming of robots in ROS: starting and stopping ROS nodes, changing node parameters, hot code swapping and connections between nodes being made at run-time. We illustrated the need for these requirements through an example. 

We have shown that the solution we propose fulfills these four requirements for ROS nodes created in PhaROS.

Moreover, we found that developing robot behaviors in the LRP language was quite straightforward. The straightforward use of state machines avoids the programmer losing focus on the task at hand due to setup or technical issues intervening. 

Also, the \ct{RobulabBridge} class made LRP code easy to write and understand, as it abstracts the robot resources and provides access through a rich API.
Finally, we consider that the solution of LRP through PhaROS is an effective approach for enabling live programming for ROS applications.

\bibliographystyle{abbrv}
\bibliography{biblioCoars}

\begin{thebibliography}{1}

\bibitem{Berg13a}
A.~Bergel, D.~Cassou, S.~Ducasse, and J.~Laval.
\newblock {\em Deep Into Pharo}.
\newblock Square Bracket Associates, 2013.

\bibitem{pharosWeb}
S.~Bragagnolo, L.~Fabresse, J.~Laval, P.~Estef{\'o}, and N.~Bouraqadi.
\newblock Pharos: a ros client for the pharo language.
\newblock http://car.mines-douai.fr/category/pharos/, 2014.

\bibitem{lrpdef}
J.~Fabry and M.~Campusano.
\newblock Live robot programming.
\newblock In A.~Bazzan and K.~Pichara, editors, {\em Advances in Artificial
  Intelligence - IBERAMIA 2014}, number 8864 in Lecture Notes in Computer
  Science. Springer-Verlag, 2014.
\newblock To Appear.

\bibitem{loetzsch06xabsl}
M.~L\"otzsch, M.~Risler, and M.~J\"ungel.
\newblock {XABSL} - {A} pragmatic approach to behavior engineering.
\newblock In {\em Proceedings of IEEE/RSJ International Conference of
  Intelligent Robots and Systems (IROS)}, pages 5124--5129, Beijing, China,
  2006.

\bibitem{pharoByExample1}
O.~Nierstrasz, S.~Ducasse, and D.~Pollet.
\newblock {\em Pharo by Example}.
\newblock Square Bracket Associates, July 2010.

\bibitem{rosICRA09}
M.~Quigley, K.~Conley, B.~P. Gerkey, J.~Faust, T.~Foote, J.~Leibs, R.~Wheeler,
  and A.~Y. Ng.
\newblock Ros: an open-source robot operating system.
\newblock In {\em ICRA Workshop on Open Source Software}, 2009.

\bibitem{tanimoto90}
S.~Tanimoto.
\newblock {VIVA}: A visual language for image processing.
\newblock {\em Journal of Visual Languages \& Computing}, 1(2):127--139, June
  1990.
\newblock http://dx.doi.org/10.1016/S1045-926X(05)80012-6.

\bibitem{kouretes12}
A.~Topalidou-Kyniazopoulou, N.~I. Spanoudakis, and M.~G. Lagoudakis.
\newblock A case tool for robot behavior development.
\newblock In X.~Chen, P.~Stone, L.~Sucar, and T.~Zant, editors, {\em RoboCup
  2012: Robot Soccer World Cup XVI}, volume 7500 of {\em Lecture Notes in
  Computer Science}, pages 225--236. Springer Berlin Heidelberg, 2013.

\end{thebibliography}

\end{document}